\documentclass[aps,prl,preprint,groupedaddress]{revtex4}

\usepackage{graphicx}

\def\barT{\overline T}
\def\barU{\overline U}
\def\barH{\overline H}
\def\barPsi{\overline\Psi}
\def\PRC#1#2#3{Phys. Rev. {\bf C#1}, #2 (#3)}
\def\PRD#1#2#3{Phys. Rev. {\bf D#1}, #2 (#3)}
\def\NPB#1#2#3{Nucl. Phys. {\bf B#1}, #2 (#3)}

\def\PLB#1#2#3{Phys. Lett. {\bf B#1}, #2 (#3)}
\def\PRL#1#2#3{Phys. Rev. Lett. {\bf #1}, #2 (#3)}
\def\PRep#1#2#3{Phys. Rep. {\bf #1}, #2 (#3)}
\def\AJ#1#2#3{Astrophys. J. {\bf #1}, #2 (#3)}

\begin{document}

\preprint{
OCHA-PP-213
}

\title{
Simple description of neutrinos in SU(5) 
}


\author{
Noriyuki Oshimo
}
\affiliation{
Institute of Humanities and Sciences {\rm and} Department of Physics \\
Ochanomizu University, Tokyo, 112-8610, Japan
}


\date{\today}

\begin{abstract}
     We show that experimental results for the masses 
and mixing of the neutrinos can be understood naturally by a simple 
grand unification model of SU(5) coupled to $N=1$ supergravity.  
No right-handed neutrinos are included.  
The left-handed neutrinos receive Majorana masses through the couplings 
with a Higgs boson of symmetric $\bf 15$ representation.   
Introducing $\overline{\bf 45}$ representation is optional for 
describing the masses of down-type quarks and charged leptons.  
\end{abstract}


\maketitle


     Accumulating experimental results for solar \cite{sol} and 
atmospheric \cite{atm} neutrino oscillations suggest that the neutrinos 
have extremely small but non-vanishing masses.  
Some extension of the standard model (SM) must be necessary.   
In addition, large angles for the generation mixing of leptons have been 
observed, contrary to the small angles for the quark generation mixing.   
Combined with experimental results for the neutrino oscillations 
by nuclear reactors~\cite{reactor}, detailed information on the 
lepton generation mixing has been obtained.  
How should the SM be extended?   

     For physics beyond the SM, irrespectively of the problem on 
the neutrinos, grand unified theories (GUTs) coupled to supersymmetry 
are very plausible from various theoretical viewpoints.   
On the other hand, one of the most popular scenarios for incorporating 
the neutrino masses is to introduce right-handed neutrinos with large 
Majorana masses \cite{gell-mann}.  
The huge mass differences between the neutrinos and the 
charged leptons could then be explained naturally.  
This scenario is embodied in the GUT models whose gauge group   
is SO(10) or a larger one.   
Unfortunately, the minimal GUT group of SU(5) 
does not contain a representation for right-handed neutrinos, 
unless ad hoc SU(5)-singlet fields are introduced.  
Therefore, SO(10) GUT models might be now considered as most 
promising for the extension of the SM.   

     One obstacle, however, confronts the GUT models 
based on SO(10) or a larger group.   
The generation mixing for the quarks and that for the leptons are described 
respectively by the Cabibbo-Kobayashi-Maskawa (CKM) matrix and the 
Maki-Nakagawa-Sakata (MNS) matrix.  
These matrices are generically related to each other in 
those models.   
The observed difference between the CKM matrix and the MNS matrix  
is not trivially understood.  
Various approaches thus have been tried to explain the 
difference \cite{oshimo}, often contrived schemes being invoked.  
The SO(10) GUT models also suffer from high dimensions of Higgs boson 
representations.    
It is a complicated procedure to set, among their various fields decomposed 
under SU(3)$\times$SU(2)$\times$U(1), 
two with quantum numbers (1,2,$-1/2$) and (1,2,1/2) 
light while the others heavy for having consistent phenomenology.   

     In this letter we present a simple GUT model which can describe 
naturally the masses and mixing of the neutrinos.   
The gauge group is the minimal SU(5) and supersymmetry is 
imposed by coupling the model to $N=1$ supergravity.   
The right-handed neutrinos are not included.  
Within the framework of SU(2)$\times$U(1) electroweak theory, 
the left-handed neutrinos could have small Majorana masses 
without fine-tuning, if an SU(2)-triplet Higgs boson is 
appropriately incorporated \cite{bajc}.   
This mechanism is naturally embedded in our model.  
The CKM matrix and the MNS matrix become independent of each other.  
Their observed difference is merely due to different values for free 
parameters, as so are the mass differences among the quarks and the leptons.  

\begin{table}
\caption{
The SU(5) quantum numbers of the superfields.    
The suffixes $a$, $b$, and $c$ denote the group indices.  
The generation index is represented by $i$ $(=1-3)$.    
\label{particles}
}
\begin{ruledtabular}
\begin{tabular}{ccccc}
$\Phi^a_b$   & $T^{ab}$     & $\barT_{ab}$      & $H^a$     & $\barH_a$ \\
$\bf 24$ & $\bf 15$ & $\overline{\bf 15}$ & $\bf 5$ & $\overline{\bf 5}$ \\
\hline
$\Psi_i^{ab}$ & $\barPsi_{ia}$ & $\bigl(U^{ab}_c\bigr)$ & 
                                       $\bigl(\barU_{ab}^c\bigr)$ & \\ 
$\bf 10$ & $\overline{\bf 5}$ & $\bf 45$ & $\overline{\bf 45}$ & \\
\end{tabular}
\end{ruledtabular}
\end{table}

     The model consists of Higgs superfields $\Phi$, $T$, $\barT$, $H$, 
and $\barH$, and matter superfields $\Psi_i$ and $\barPsi_i$, 
with $i$ being the generation index.  
Their quantum numbers are shown in Table \ref{particles}.   
In addition to the particle contents of the minimal SU(5) model, 
superfields of symmetric $\bf 15$ and $\overline{\bf 15}$ representations 
are introduced.  
This $\bf 15$ representation can couple to 
$\overline{\bf 5}\times\overline{\bf 5}$ of matters, leading to Majorana 
masses for the left-handed neutrinos at the electroweak energy scale.  
The complex conjugate $\overline{\bf 15}$ representation is needed 
in order to render $T$ very heavy, as well as anomaly free.  
These particle contents alone, however, may encounter a difficulty 
in describing masses of some down-type quarks or charged leptons, which is 
a common problem for ordinary minimal SU(5) models~\cite{langacker}.  
One of its solutions is to introduce an additional Higgs superfield $\barU$ 
of $\overline{\bf 45}$ representation, which couples to the matters 
$\bf 10\times\overline{5}$.  
In this case, the complex conjugate $\bf 45$ representation is also 
included for having a mass term and anomaly cancellation.    
This option is discussed afterward.  

     The superpotential of the model is given by the sum of two sectors,  
\begin{eqnarray}
    W_H &=& M_H{\barH}H + M_T {\rm Tr}[{\barT}T] 
          + \frac{1}{2}M_\Phi {\rm Tr}[\Phi^2] 
                            \nonumber \\
      &+&  \lambda_{H\Phi} {\barH}\Phi H 
          + \lambda_{T\Phi} {\rm Tr}[{\barT}\Phi T] 
          + \frac{1}{3}\lambda_\Phi{\rm Tr}[\Phi^3] 
                            \nonumber \\
      &+&  \lambda_{HT}H{\barT}H + {\overline{\lambda}_{HT}}{\barH}T{\barH},  
\label{superpotentialH}  \\
   W_M &=& \Gamma_{ij}^u\epsilon H\Psi_i\Psi_j  
         + \Gamma_{ij}^{de}{\barH}\Psi_i{\barPsi_j}   
         + \Gamma_{ij}^\nu {\barPsi_i}T{\barPsi_j},   
\label{superpotentialM} 
\end{eqnarray}
where $\epsilon$ denotes the totally antisymmetric tensor of rank five.  
Contraction of SU(5) group indices is understood.  
Involved are all the renormalizable terms consistent with 
SU(5) and $R$ parity.  
The mass parameters $M_H$, $M_T$, and $M_\Phi$ have values of the 
order of the GUT energy scale $M_X$, which is given typically by 
$M_X\sim 10^{16}$ GeV.  
The superpotential $W_H$ contains only Higgs superfields and 
determines the vacuum.  
The vacuum expectation value (VEV) of $\Phi$ is responsible for 
breakdown of SU(5), and those of $H$ and $\barH$ are to break 
SU(2)$\times$U(1).   
The VEV of $T$ induces masses of the left-handed neutrinos.  
The masses of the quarks and leptons are generated by the 
superpotential $W_M$.   
The coefficients $\Gamma^u$ and $\Gamma^\nu$ are symmetric 
for generation indices.  

     The SU(5) gauge symmetry is broken at the vacuum.  
Although the superpotential $W_H$ yields three 
degenerated vacua under global supersymmetry, 
this degeneracy is lifted by $N=1$ supergravity~\cite{weinberg}.  
The scalar component of the adjoint representation $\Phi$ could have a VEV 
$\langle \Phi \rangle= {\rm diag}(1, 1, 1, -3/2,-3/2)v_\Phi$,  
where $v_\Phi$ is given by 
$v_\Phi\simeq 2{\rm Re}(M_\Phi)/{\rm Re}(\lambda_\Phi)$ and related to 
the $X$ and $Y$ boson masses as $M_X^2=M_Y^2=(25/8)g_5^2v_\Phi^2$.  
The VEVs of the other scalar fields vanish.  
Then, the gauge symmetry after the breaking of SU(5) is 
SU(3)$\times$SU(2)$\times$U(1) of the SM.  

     Below the GUT energy scale, the vacuum is prescribed by the 
scalar potential which consists of the SU(3)-singlet components 
for $H$, $\barH$, $T$, $\barT$, and $\Phi$.  
The relevant part of the superpotential $W_H$ is written as 
\begin{eqnarray}
    W_H &=& -m_H H_1\epsilon H_2 + m_T {\rm Tr}[{\barT}T]  
          + \frac{1}{2}m_\Phi {\rm Tr}[\Phi^2] 
                            \nonumber \\
      &+&  \lambda_{\phi 1} (\epsilon H_1)^T\Phi H_2 
          + \lambda_{\phi 2} {\rm Tr}[{\barT}\Phi T] 
          + \frac{1}{3}\lambda_{\phi 3}{\rm Tr}[\Phi^3] 
                            \nonumber \\
    &+& \overline{\lambda}(\epsilon H_1)^T T\epsilon H_1 
          + \lambda (H_2)^T{\barT}H_2, 
\label{superpotentialH2}  
\end{eqnarray}
where $\epsilon$ stands for the totally antisymmetric tensor of rank two.    
The SU(2)-doublet components of $H$ and $\barH$ are expressed by 
$H_2$ and $H_1$ as $H_2=(H^4, H^5)$ and $H_1=(-\barH_5, \barH_4)$.  
The SU(2)-triplet components of $\Phi$, $T$, and $\barT$ 
are denoted by the same symbols.   
At the GUT energy scale, the coefficients have the values  
$m_H=M_H-(3/2)\lambda_{H\Phi}v_\Phi$, $m_T=M_T-(3/2)\lambda_{T\Phi}v_\Phi$, 
$m_\Phi=M_\Phi-3\lambda_{\Phi}v_\Phi$, 
$\lambda_{\phi 1}=\lambda_{H\Phi}$, 
$\lambda_{\phi 2}=\lambda_{T\Phi}$, 
$\lambda_{\phi 3}=\lambda_{\Phi}$, 
$\lambda=\lambda_{HT}$, and $\overline{\lambda}=\overline{\lambda}_{HT}$.  
We assume that the magnitude of $m_H$ becomes of the order of 
the electroweak energy scale $M_W$ by fine-tuning or some other 
mechanism, while the SU(3)-triplet components of $H$ and $\barH$ 
remain of the order of $M_X$.  
The mass parameters $m_T$ and $m_\Phi$ also have a natural magnitude of 
the order of $M_X$.  

     The superpotential for the quark and lepton masses is written as 
\begin{eqnarray}
W_M &=& \eta_d^{ij} H_1\epsilon Q^iD^{cj} + \eta_u^{ij}H_2\epsilon Q^iU^{cj}
 + \eta_e^{ij} H_1\epsilon L^iE^{cj} 
           \nonumber \\
 &+&  \frac{1}{2}\kappa^{ij}(\epsilon L^i)^T T\epsilon L^j,  
\label{superpotentialM2} 
\end{eqnarray}
where $Q^i$, $U^{ci}$, $D^{ci}$, $L^i$, and $E^{ci}$ stand for the 
superfields for the quarks and leptons in a self-explanatory notation.   
There exist possible Majorana mass terms for the left-handed neutrinos.  
At the GUT energy scale, the coefficients have the values 
$\eta_d=-\Gamma^{de}/\sqrt{2}$, $\eta_u=4\Gamma^u$, 
$\eta_e=-(\Gamma^{de})^T/\sqrt{2}$, and $\kappa=2\Gamma^\nu$.  
The physical parameters for the coefficients are given by 
the CKM matrix $V_{CKM}$, the MNS matrix $V_{MNS}$, and the diagonalized 
eigenvalue matrices $\eta_d^D$, $\eta_u^D$, $\eta_e^D$, and $\kappa^D$.  

     The scalar potential for $H_1$ and $H_2$ is mostly the same as 
the minimal supersymmetric standard model, since the 
superfields $T$, $\barT$, and $\Phi$ have masses of the order of $M_X$.  
In the supersymmetry-soft-breaking Lagrangian, the Higgs bosons 
$H_1$ and $H_2$ have positive masses-squared of the order of $M_W$ 
at the GUT energy scale.   
However, owing to a large value of $\eta_u^{ij}$ corresponding to 
the $t$ quark mass, the mass-squared of $H_2$ is driven 
negative at the electroweak energy scale through quantum corrections.  
Non-vanishing VEVs are then induced for those Higgs bosons and 
SU(2)$\times$U(1) is broken down to U(1)$_{\rm EM}$ \cite{nilles}.    
The VEVs of $H_1$ and $H_2$ are expressed by 
$\langle H_1\rangle =(v_1/\sqrt{2}, 0)$ and 
$\langle H_2\rangle =(0, v_2/\sqrt{2})$, 
where $v_1$ and $v_2$ are related to the $W$-boson mass as 
$M_W^2=(1/4)g_2^2(v_1^2+v_2^2)$.  

     We first show that the electroweak symmetry breaking induces 
non-vanishing VEVs for $T$ and $\barT$.  
Assuming that U(1)$_{\rm EM}$ symmetry is not broken, 
we can put the VEVs of these SU(2)-triplet fields at 
$\langle T \rangle={\rm diag}(0,v_T/\sqrt{2})$ and 
$\langle \barT \rangle={\rm diag}(0,v_{\barT}/\sqrt{2})$.  
Since the scalar components of $T$, $\barT$, and $\Phi$ have large 
positive masses-squared of the order of $M_X$ through the $F$-term 
scalar potential, the magnitudes of $v_T$ and $v_{\barT}$, as well as 
the VEV of $\Phi$, must be extremely smaller than $v_1$ and $v_2$, 
albeit non-vanishing.  
The contribution of $v_T$ and $v_{\barT}$ to 
the VEV of the scalar potential is given by 
\begin{eqnarray}
V &=& \frac{1}{2}|m_T|^2\left(|v_T|^2+|v_{\barT}|^2\right) 
           \nonumber \\
 &+& \frac{1}{\sqrt{2}}{\rm Re}\left(\lambda m_T^*v_2^2v_T^*\right) 
  +\frac{1}{\sqrt{2}}
{\rm Re}\left(\overline{\lambda} m_T^*v_1^2v_{\barT}^*\right),   
\label{potential}
\end{eqnarray}
where we have neglected the terms whose magnitudes should be smaller than 
the order of $(M_W)^4$.  
The non-negligible terms all arise from the $F$-term scalar potential.  
The values for $v_T$ and $v_{\barT}$ become non-vanishing at the vacuum, 
which are given by  
\begin{eqnarray}
 |v_T| = \left|\frac{\lambda v_2^2}{\sqrt{2}m_T}\right|, 
 |v_{\barT}| = \left|\frac{\overline{\lambda}v_1^2}{\sqrt{2}m_T}\right|.  
\end{eqnarray}
For $m_T\sim 10^{14}$ GeV and $v_1$, $v_2\sim 10^{2}$ GeV, 
the VEVs $v_T$ and $v_{\barT}$ are of the order of $10^{-1}$ eV.  
The left-handed neutrinos can have the masses which are of 
the correct order of magnitude.   
It should be noted that the appropriate value for $m_T$ is 
smaller than $M_X$ by one or two orders of magnitude.  

     We now turn to the discussion of the CKM and MNS matrices and 
the eigenvalue matrices for the Higgs coupling coefficients.  
The values of these matrices depend on the energy scale of physics, 
which are governed by the renormalization group equations for 
the coefficient matrices $\eta_d$, $\eta_u$, $\eta_e$, and $\kappa$ 
appearing in Eq. (\ref{superpotentialM2}).  
Between the energy scales $M_X$ and $M_W$, 
the renormalization group equation of $\kappa$ is given at one-loop 
level by 
\begin{eqnarray}
\mu\frac{d\kappa}{d\mu} &=& \frac{1}{16\pi^2}
 \biggl\{\left(5g_2^2+\frac{9}{5}g_1^2 
          + \frac{1}{2}{\rm Tr}[\kappa^\dagger\kappa]\right)\kappa 
           \nonumber \\
& &  +\eta_e\eta_e^\dagger\kappa + \kappa\eta_e\eta_e^\dagger \biggr\},
\label{renomalization}
\end{eqnarray}
where $g_1$ denotes the normalized U(1) gauge coupling constant.   
We have taken the generation basis in which $\eta_e$ is diagonal.  
For calculations of self-energies and vertex corrections, the 
contributions from the superfields $T$ and $\barT$ have been neglected.   
The renormalization group equations for $\eta_d$, $\eta_u$, and $\eta_e$ 
are known in the literature.  
Experimentally measured quantities are the CKM and MNS matrices 
at the electroweak energy scale.   
The eigenvalue matrices $\eta_d^D$, $\eta_u^D$, $\eta_e^D$, 
and $\kappa^D$ at this energy scale could also be determined, 
provided that the ratios 
of $v_1$ to $v_2$ and to $v_T$ are given.  

     The superpotential $W_M$ in Eq. (\ref{superpotentialM}) can  
accommodate any values for the CKM and MNS matrices.  
Any masses for the up-type quarks and neutrinos can also be realized.  
Given the experimentally determined values for $V_{CKM}$, $V_{MNS}$, 
$\eta_u^D$, $\eta_d^D$, $\eta_e^D$, and $\kappa^D$,  
the values which these matrices should have at the GUT energy scale 
are obtained by the renormalization group equations.  
For expressing the coefficient matrices, we take 
the generation basis in which $\eta_d$ and $\eta_e$ are diagonal.   
The required values for $V_{CKM}$, $V_{MNS}$, $\eta_u^D$, and $\kappa^D$ 
are fulfilled by taking the coefficients as  
\begin{eqnarray}
\Gamma^u &=& \frac{1}{4}(V_{CKM})^T \eta_u^D V_{CKM},  \\  
\Gamma^\nu &=& \frac{1}{2}(V_{MNS})^T \kappa^D V_{MNS}.  
\label{coefficient-nu} 
\end{eqnarray}
These equations can always be satisfied, since the coefficients 
$\Gamma^u$, $\Gamma^{de}$, and $\Gamma^\nu$ are independent mutually.   

     A problem arises from the masses of down-type quarks and charged 
leptons.   
According to $W_M$ in Eq.~(\ref{superpotentialM}), the coefficient 
matrices $\eta_d$ and $\eta_e$ should have the same eigenvalues 
at the GUT energy scale, 
\begin{eqnarray}
\Gamma^{de} = -\sqrt{2}\eta_d^D = -\sqrt{2}\eta_e^D.   
\label{de-relation}
\end{eqnarray}
However, the experimental results do not naively lead to these 
coincidences, except for the $b$ quark and the $\tau$ lepton.  
As an example, let us take $m_d=4.0$ MeV, $m_s=1.1\times 10^2$ MeV, and 
$m_b=3.1$ GeV for the down-type quark masses, and 
$m_e=4.9\times 10^{-1}$ MeV, $m_\mu=1.0\times 10^2$ MeV, and 
$m_\tau=1.7$ GeV for the charged lepton masses at the electroweak 
energy scale.  
The ratio of the VEVs is put at $v_2/v_1=30$.  
Then, at the GUT energy scale, the eigenvalue matrices are given by 
$\eta_d^D=(2.1\times 10^{-4}, 5.8 \times 10^{-3}, 2.0 \times 10^{-1})$ and 
$\eta_e^D=(6.0\times 10^{-5}, 1.2 \times 10^{-2}, 2.2 \times 10^{-1})$.  
For the first two generations, there appear differences by a factor of 
three or two.   
This difficulty could be relieved if the difference of a few orders of 
magnitude between $M_X$ and $m_T$ is correctly taken into account in the 
analysis by renormalization group equations.   
Furthermore, the determination of the light quark masses by experimental  
results is very difficult.  
Therefore, the above mentioned discrepancy may not be serious 
for the present model.  

     The problem of the down-type quark and charged lepton masses 
could eventually be solved by including the Higgs superfields $U$ and 
$\barU$ of $\bf 45$ and $\overline{\bf 45}$.  
Then, the superpotentials in Eqs. (\ref{superpotentialH}) and 
(\ref{superpotentialM}) have additional terms 
\begin{eqnarray}
 W_H &=&  M_U\barU U + \lambda_{U\Phi}\barU\Phi U + \lambda_{HU}H\barU\Phi 
                    \nonumber \\ 
 & & + \overline{\lambda}_{HU}\barH U\Phi 
     + \lambda_{TU}\barU T\barU + \overline{\lambda}_{TU}U\barT U, 
                  \\
 W_M &=&  \overline{\Gamma}_{ij}^{u}\epsilon U\Psi_i\Psi_j +  
      \overline{\Gamma}_{ij}^{de}\barU\Psi_i\barPsi_j. 
\end{eqnarray}
A new source of masses for the down-type quarks and charged leptons 
is involved, which invalidates the relation in Eq.~(\ref{de-relation}).  
The up-type quark masses also receive new contributions.  
The coefficient $\overline{\Gamma}^u$ is antisymmetric 
for generation indices.  
The SU(2)-doublet components of $U$ and $\barU$ for the Higgs superfields 
are formed as $(2\sqrt{6}/3)(-U_5^{45}, -U_4^{54})$ and 
$(2\sqrt{6}/3)(\barU^4_{54}, -\barU^5_{45})$, 
with U(1) hypercharges 1/2 and $-1/2$, respectively.  
The factor $2\sqrt{6}/3$ is multiplied for the normalization.  

     The Higgs superfields $H_1$ and $H_2$ in Eq.~(\ref{superpotentialM2}) 
are composed of the SU(2)-doublet superfields $H_2^5$, 
$H_1^{\overline{5}}$, $H_2^{45}$, and $H_1^{\overline{45}}$ in respectively 
$\bf 5$, $\overline{\bf 5}$, $\bf 45$, and $\overline{\bf 45}$.   
After breakdown of SU(5), they are mixed by the the term 
\begin{eqnarray}
 W &=& \left( H_1^{\overline{5}} H_1^{\overline{45}} \right) M\epsilon  
   \left(
    \begin{array}{c}
  H_2^5   \\
  H_2^{45} 
        \end{array}
            \right),   \\ 
  M &=&       \left(
        \begin{array}{cc}
  -M_H+\frac{3}{2}\lambda_{H\Phi}v_\Phi & 
             -\frac{15}{4\sqrt{6}}\overline{\lambda}_{HU}v_\Phi  \\
  -\frac{15}{4\sqrt{6}}\lambda_{HU}v_\Phi  & 
             M_U-\frac{19}{16}\lambda_{U\Phi}v_\Phi  
        \end{array}
        \right).    \nonumber  
\end{eqnarray}
Assuming an approximate equation for the elements of the mass matrix as 
$M_{11}M_{22}-M_{12}M_{21}\sim M_WM_X$, a pair of linear combinations of  
the SU(2)-doublet superfields have a small mass term of the order of $M_W$.  
Electroweak symmetry breaking is due to this pair of Higgs superfields, 
which are expressed by 
\begin{eqnarray}
 H_1 &=&  (C_1^\dagger)_{11}H_1^{\overline{5}} 
               + (C_1^\dagger)_{12}H_1^{\overline{45}},  \\ 
 H_2 &=&  (C_2^\dagger)_{11}H_2^5 + (C_2^\dagger)_{12}H_2^{45}.    
\end{eqnarray}
Here, $C_1$ and $C_2$ stand for the unitary matrices which diagonalize 
the mass matrix by $(C_1)^TMC_2$.  
The other pair of SU(2)-doublet superfields have naturally a large 
mass term of the order of $M_X$.  
Decoupling from theory below the GUT energy scale, these superfields 
do not affect flavor-changing neutral current processes at the 
electroweak energy scale nor energy evolutions of the gauge coupling 
constants for SU(3), SU(2), and U(1).  

     The masses of the down-type quarks 
and charged leptons can now be accommodated.  
At the GUT energy scale, the coefficient matrices appearing in 
Eq. (\ref{superpotentialM2}) are given by 
$\eta_d=-(C_1)_{11}\Gamma^{de}/\sqrt{2}
          -(C_1)_{21}\overline{\Gamma}^{de}/2\sqrt{3}$, 
$\eta_e=-(C_1)_{11}(\Gamma^{de})^T/\sqrt{2}
          +3(C_1)_{21}(\overline{\Gamma}^{de})^T/2\sqrt{3}$, and 
$\eta_u=4(C_2)_{11}\Gamma^u -4(C_2)_{21}\overline{\Gamma}^u/\sqrt{6}$.  
Taking the generation basis in which $\eta_d$ and $\eta_e$ are 
diagonal, any values for the down-type quark and charged lepton masses 
are described in this model by imposing the conditions 
\begin{eqnarray}
    \Gamma^{de} &=& -\frac{1}{2\sqrt{2}(C_1)_{11}}
                           \left(3\eta_d^D+\eta_e^D\right),  \\
    \overline{\Gamma}^{de} &=& -\frac{\sqrt{3}}{2(C_1)_{21}}
                           \left(\eta_d^D-\eta_e^D\right).  
\end{eqnarray}
The coefficients for the up-type quark masses should be taken as 
\begin{eqnarray}
    \Gamma^u &=& \frac{1}{8(C_2)_{11}}
       \Bigl\{(V_{CKM})^T\eta_u^D(U_R^u)^T    \nonumber  \\
         & &  + U_R^u\eta_u^DV_{CKM}\Bigr\},  \\ 
    \overline{\Gamma}^u &=& -\frac{\sqrt{6}}{8(C_2)_{21}}
       \Bigl\{(V_{CKM})^T\eta_u^D(U_R^u)^T   \nonumber  \\
         & & - U_R^u\eta_u^DV_{CKM}\Bigr\},  
\end{eqnarray}
where $U_R^u$ is an arbitrary unitary matrix.  
For the neutrinos, the condition on $\Gamma^\nu$ is 
not altered and given by Eq.~(\ref{coefficient-nu}).   

     In summary, we have shown that the experimental results for 
the neutrinos are understood naturally in a GUT model based on SU(5) gauge 
group and $N=1$ supergravity.  
Only by incorporating symmetric $\bf 15$ and $\overline{\bf 15}$ 
representations for new Higgs superfields, the neutrinos can have, 
without fine-tuning, the masses which are observed.  
The CKM matrix and the MNS matrix become independent of each other.  
Their difference is considered a reasonable consequence of  
the fact that the relevant parameter values are freely given at 
the GUT energy scale.   
Inclusion of $\bf 45$ and $\overline{\bf 45}$ representations is possible 
to accommodate the down-type quark and charged lepton masses.  

\begin{acknowledgments}
     This work is supported in part by the Grant-in-Aid for 
Scientific Research on Priority Areas (No. 14039204) from the 
Ministry of Education, Science and Culture, Japan.   
\end{acknowledgments}


\begin{references}
\bibitem{sol}
   B.T. Cleveland et al., \AJ{496}{505}{1998};  \\
   W. Hampel et al. (GALLEX Collaboration), \PLB{447}{127}{1999};  \\  
   J.N. Abdurashitov et al. (SAGE Collaboration), \PRC{60}{055801}{1999};  \\
   M. Altmann et al. (GNO Collaboration), \PLB{490}{16}{2000};  \\
   S. Fukuda et al. (Super-Kamiokande Collaboration), \PRL{86}{5656}{2001}; \\
   Q.R. Ahmad et al. (SNO Collaboration), \PRL{89}{011301}{2002}.   
\bibitem{atm}
     S. Fukuda et al. (Super-Kamiokande Collaboration), \PRL{85}{3999}{2000}.
\bibitem{reactor} 
    M. Apollonio et al., \PLB{466}{415}{1999};  \\
    K. Eguchi et al. (KamLAND Collaboration), \PRL{90}{021802}{2003}.
\bibitem{gell-mann}
    M. Gell-Mann, P. Ramond, and R. Slansky, in Proceedings of the 
    Supergravity Stony Brook Workshop, edited by P. Van Niewenhuizen 
    and D. Freeman (North-Holland, 1979);  \\
    T. Yanagida, in Proceedings of the Workshop on Unified Theories 
    and Baryon Number in the Universe, edited by A. Sawada and 
    A. Sugamoto (KEK Report No. 79-18, 1979).  
\bibitem{oshimo}
     See, e.g., N. Oshimo, \PRD{66}{095010}{2002}; 
      \NPB{668}{258}{2003}, and references therein.
\bibitem{bajc}
   See, e.g., B. Bajc, G. Senjanovi\'c, and F. Vissani, 
                                    \PRL{90}{051802}{2003};  \\ 
   H.S. Goh, R.N. Mohapatra, and S.-P. Ng, \PLB{570}{215}{2003}.   
\bibitem{langacker}
    For a review, see, e.g., P. Langacker, \PRep{72}{185}{1981}.  
\bibitem{weinberg}
     S. Weinberg, \PRL{48}{1776}{1982};  \\
     A.H. Chamseddine, R. Arnowitt, and P. Nath, \PRL{49}{970}{1982}.
\bibitem{nilles}
     For a review, see, e.g., H.P. Nilles, \PRep{110}{1}{1984}.  

\end{references}
\end{document}